\documentclass[12pt,preprint]{aastex}
\shorttitle{Multicolor photometry of W Virginis}
\shortauthors{Templeton \& Henden}
\received{2007 July 31}
\begin{document}
\title{Multicolor photometry of the Type II Cepheid prototype W Virginis}
\author{M.R. Templeton \& A.A. Henden}
\affil{American Association of Variable Star Observers, 49 Bay State Road, Cambridge, MA 02138}

\begin{abstract}
We present the results of recent long-term $BVRcIc$ photometric monitoring of 
the type II Cepheid prototype W Virginis.  These new observations, made during
the 2006 and 2007 observing season, represent the longest homogeneous, 
multicolor light curve of W Vir to date.  The $BVRcIc$ light and color
curves show conclusively that W Vir exhibits modest but detectable 
cycle-to-cycle variations, the cause of which appears to be multiperiodicity
rather than nonlinearity.  We combined our $V$-band data with the five
available years of {\it ASAS-3} $V$-band photometry to obtain a 6.5-year 
light curve that we then analyzed to obtain the pulsation spectrum of W Vir.
We find a best-fit
period $P_{0} = 17.27134$ days; along with this period and the integer-ratio
harmonics $P_{0}/2$ through $P_{0}/5$ inclusive, we clearly detect two 
additional periods, $P_{1}$ and $P_{\rm{low}}$ that are close to but not
exactly $2P_{0}/3$ and $2P_{0}$, respectively.  The former, $P_{1} = 11.52562$
days, we interpret to be the first overtone mode; the latter, $P_{\rm{low}} =
34.59760$ days is close to the beat period of 
$((P_{1})^{-1} - (P_{0})^{-1})^{-1}$, as well as to the value of $2P_{0}$.  We
interpret the previously reported but thus far unconfirmed descriptions of
alternating minima as manifestations of this multiperiodicity.  Finally,
we use the period derived from the $V$-band light curve to define a new
ephemeris: ${\rm HJD_{max}} = 2452758.172 + 17.27134 \times E$.  We compiled
an $(O-C)$ diagram spanning 75 years from 1932 to 2007 using a variety of 
published
photometric data and AAVSO visual observations, and derived a period change 
term for the ephemeris equal to $-9.9 \times 10^{-7} E^{2}$, indicating a 
period decrease.

\end{abstract}


\section{Introduction}

W Virginis (AAVSO 1320-02: J2000 RA: 13 26 1.993, Dec: -03 22 43.42) is the
class prototype of the Type II Cepheid variables.  It has a period of
about 17.3 days, and is relatively bright with maxima reaching $V \sim 9.5$. 
Like many bright variables, it has become less-frequently monitored and
studied in recent years.  While several robotic telescopes monitor this and 
other objects as part of their
all-sky campaigns, very little recent work has been done to perform
intensive, multiwavelength photometry of W Vir itself.  Despite being a
cosmologically important class prototype, the behavior of W Vir is still
not fully understood.  In particular, it is known to exhibit
cycle-to-cycle variations, like all long-period Cepheids.  These
variations can only be reliably tracked with continuous, season-long
observing campaigns, and this is where the AAVSO and its members and
partner observatories represent a remarkable new opportunity for the
astronomical community to study neglected, interesting objects.

The AAVSO was asked to provide photometric support for high-resolution
spectroscopic observations of W Virginis (Wallerstein et al., in preparation)
throughout 2006, and we initiated an observing campaign that included
both amateur and professional observers, performing visual and
instrumental photometry.  These data are being used in the
interpretation of the resulting spectroscopy, but they also represent a
substantial new data set for this historic star.  In particular, nightly
$BVRcIc$ observations with the {\it Sonoita Research Observatory} 14-inch
have yielded a valuable new multicolor dataset.  When combined with
$V$-band data from the {\it All Sky Automated Survey} \citep{Pojmanski02},
our $V$-band light curve spans 6.5 years, and covers nearly every cycle 
during the observing season.  This is an important new photometric light curve
for W Vir, allowing us to study the multicolor behavior
of this fundamentally important object on years-long timescales.

In this paper, we present new, multicolor photometry for W Vir taken during 
2006 and 2007.  We present continuous and folded light and color curves, 
describe the quality and statistical properties of the {\it Sonoita} data, 
and present a comprehensive analysis of this important new data set.  In
Section 2, we describe in full the newly acquired observations, as well as
the archival data used in our analyses.  In Section 3, we describe our
analyses including derivation of the full variability spectrum, determine
a new ephemeris for this star, and present the results of our $V$-band and 
multicolor analyses.  In Section 4, we discuss the physical consequences of
our findings with particular emphasis on the unambiguous detection of the 
first overtone mode, and recommend future avenues for observational
and theoretical research on W Vir and on the Type II Cepheids in general.

\section{Observations}

Despite the astrophysical importance of W Vir, little concerted observational
work was done on this star other than long-term monitoring by visual observers,
short-term $UBV$ photometric campaigns covering a few cycles, or monitoring by 
automated systems (e.g.~{\it Hipparcos} and {\it ASAS-3}).  Because
of this, the novel observations conducted here represent a major increase in
the observational record for W Vir, and we discuss these observations in some
detail.  

\subsection{{\it Sonoita Research Observatory} Observations}
W Vir has been observed over two seasons using the 
{\it Sonoita Research Observatory} (SRO) 35-cm telescope, located in southern
Arizona.  An SBIG STL-1001E CCD camera was used with Johnson/Cousins 
$BVRcIc$ filters.  The pixel scale is 1.25 arcsec/pixel, yielding
a field of view of 21x21 arcminutes.

Each image was dark-subtracted and flatfielded using standard techniques.
All stars in each image were extracted using routines from DAOPHOT 
\citep{Stetson87}.  Aperture photometry was performed, typically using 
apertures whose diameter was equal to 4 times the full-width-half-maximum of 
the star profiles.

Since W Vir is located at galactic coordinates (319.5659, +58.3713), no 
stars of comparable magnitude lie within the field of view.  We therefore 
offset the field to the west to acquire a handful of fainter, but usable, 
comparison stars.  A total of 6 stars, ranging from V=11.0 to V=13.6, form 
an ensemble.  Using the inhomogeneous ensemble techniques similar to 
\citet{Honeycutt92}, we perform differential photometry of W Vir using 
these six stars, resulting in an average per-observation error of five 
millimagnitudes.  The Celestron C14 optical telescope assembly has
strong vignetting across the field.  While flatfielding removes essentially 
all of this vignetting, a small residual offset may remain.  We positioned 
the field exactly the same on all visits to ensure that any residual will 
just cause a systematic offset from the standard system.

All stars in the 21x21 arcminute field have been calibrated by the SRO telescope
on 42 photometric nights.  On each night, \citet{Landolt83,Landolt92}
standards were observed to obtain transformation coefficients, along with a 
single A0 star that was followed from meridian to large airmass.  These 
observations resulted in mean errors of about five millimagnitudes per 
magnitude for the comparison stars.  The stars used are given in Table 
\ref{compstars}.  Photometry of all 115 stars in the field can be found on 
the AAVSO web site.  Astrometry of these comparison stars was determined 
using the UCAC2 reference catalog \citep{Zacharias04} along with the SLALIB
astronomy utility software package \citep{Wallace94}.  Coordinates are given 
in decimal degrees, and are valid for Epoch 2006.77.  Errors are standard 
deviation of the mean.

\subsection{Other observational data}
Other major sources of data include: the {\it All-Sky Automated
Survey} \citep{Pojmanski02}, and {\it Hipparcos} photometry 
\citep{vanLeeuwen97} transformed to $V$ using the calibration of 
\citet{Harmanec98}; and the AAVSO International Database.  We also made use
of previously published observations of W Vir collected and generously
provided by L.~Berdnikov (2002, private communication); these data include
a number of $UBV$ and $BV$ data sets published in
\citet{Abt54} (observed by Whitford and Code at Mount Wilson), \citet{Eggen57},
and \citet{Arp57}.  All but the {\it ASAS-3} data were used solely in the
compilation of our $(O-C)$ diagram, though we note the early $UBV$ data of
\citet{Arp57} were also used in the transformation of the {\it Hipparcos}
data to $V$.

\subsection{$V$-band light curve}

\begin{figure}
\figurenum{1}
\label{fig-vlc}
\epsscale{0.85}
\plotone{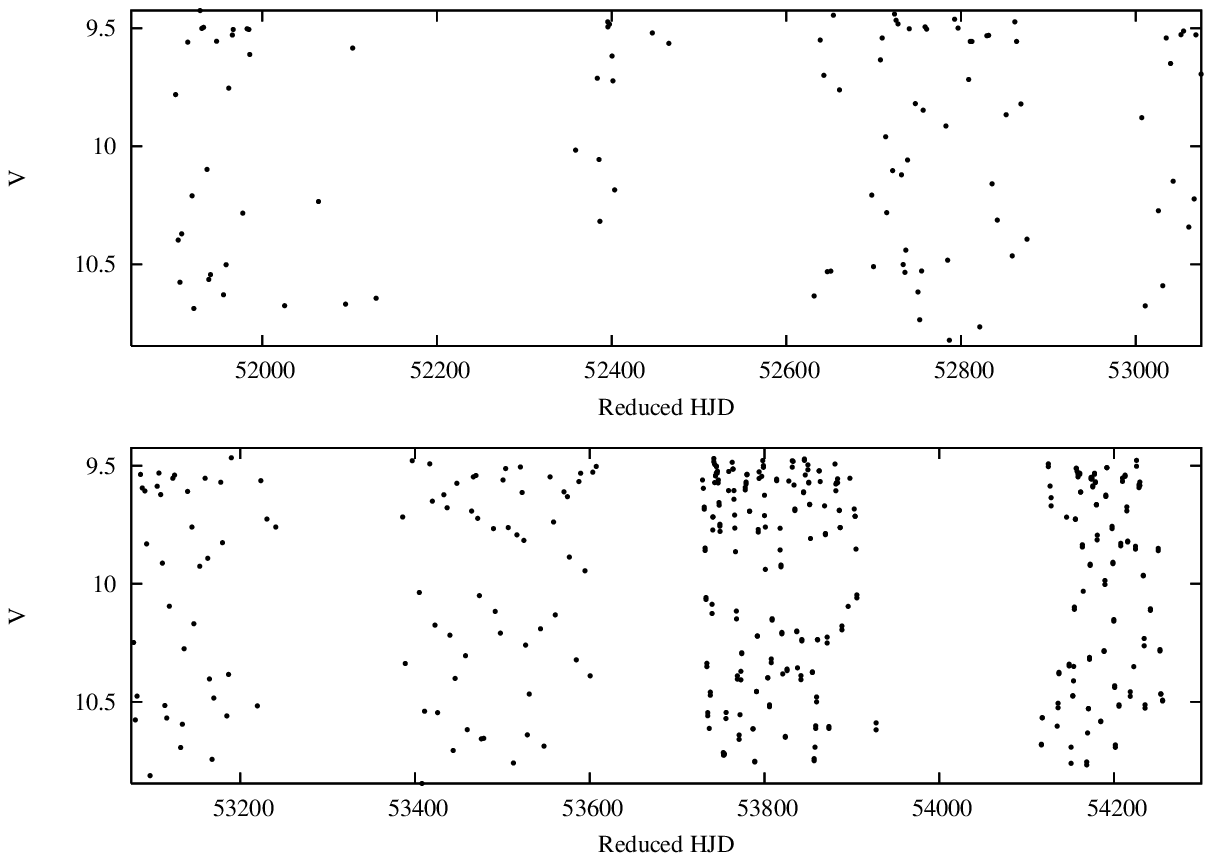}
\caption{$V$-band {\it ASAS-3 + SRO} light curve of W Virginis spanning the 
dates of December 2000 to May 2007.}
\end{figure}

We merged 6.5 years of $V$-band observations from SRO and from the archives
of the {\it ASAS-3} project \citep{Pojmanski02} to form
a long-term light curve and perform a detailed time-series analysis.  We
limited this light curve to these two data sources because: (a) both are
well-sampled, relatively homogeneous, and fully calibrated on a standard
photometric system, and (b) the rate of period change in W Vir is significant
enough to introduce smearing of the signal over much longer timescales.
The 6.5-year long {\it ASAS-3 + SRO} $V$-band light curve of
W Virginis is shown in Figure \ref{fig-vlc}.  Aside from the large annual
observing gaps due to solar conjunction, the $V$-band light curve of W Vir
is nearly continuous with at least a few (and often many) points obtained
per cycle, and several cycles obtained during each observing season.

\section{Analysis}

We have three primary goals in performing this work.  First we want to use
a homogeneous, photometrically calibrated data set to
determine the pulsation spectrum of W Virginis over a long interval and define
a new ephemeris.  Second, we want to study the long-term evolution of the 
light curve of W Virginis, and determine whether the changes are due to 
multiple periods, secular period changes, nonlinear behavior, or some 
combination of these.  Third, we want to use the shorter set of multicolor
photometry to study the multiwavelength behavior of W Virginis, to gain more
physical insight into the pulsations.  To study the periodicities and period 
evolution of W Vir, we used both Fourier and $(O-C)$ analyses on the primary 
$V$-band dataset, on the new SRO $BVRcIc$ dataset, and on data from the AAVSO 
International Database.  We will discuss each of these in turn below.

\subsection{Fourier analysis of a 6.5-year $V$-band light curve}

\begin{figure}
\figurenum{2}
\label{fig-ftclean}
\epsscale{0.75}
\plotone{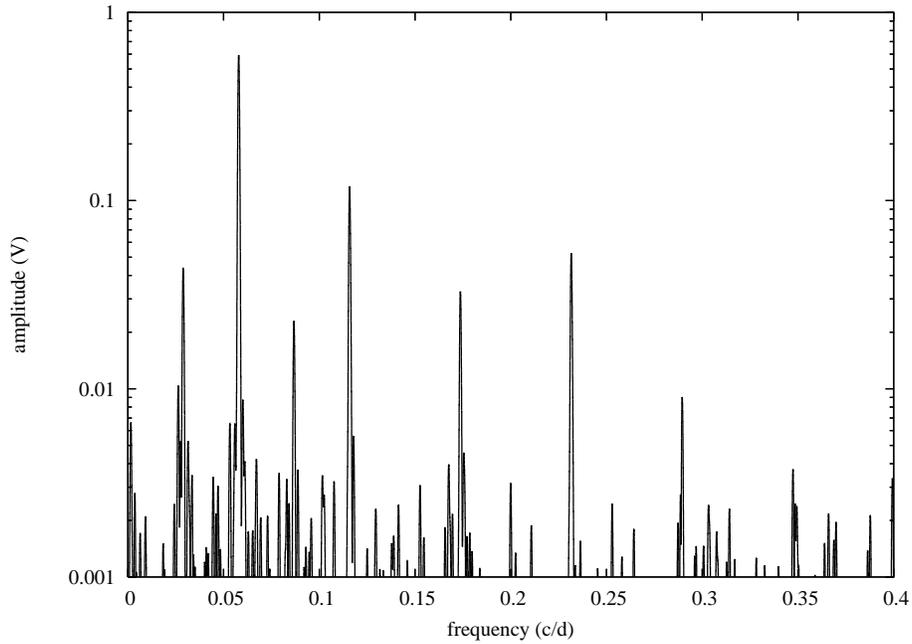}
\caption{Cleaned Fourier spectrum of the 6.5-year $V$-band light curve of
W Virginis, with amplitudes shown on a log scale for clarity.  The spectrum 
is dominated by the fundamental pulsation mode at a frequency of 0.0579 cycles
per day, and its integer harmonics.  There are two highly significant peaks
at 0.0289 and 0.0867 cycles per day; the latter is the first overtone, and
the former is a subharmonic beat frequency of the fundamental and first 
overtone modes.}
\end{figure}

First, we performed a thorough Fourier analysis of the 6.5-year $V$-band
light curve, composed of {\it ASAS-3} $V$ magnitudes and $V$ magnitudes 
obtained with SRO.  We chose not to include transformed {\it Hipparcos}
magnitudes because of the large gap between the end of {\it Hipparcos} 
observations and the start of {\it ASAS-3} observations, and because the
period changes significantly enough over the 17-year span of the longer
data set that it affects the Fourier analysis.
Our analysis was performed using a newly written implementation of the Fourier
``clean'' algorithm outlined in \citet{Roberts87} for which the Fortran 90
source code is available upon request.  For these calculations, we
tested all frequencies between 0 and 0.8 c/d with a resolution 
$\Delta f = 6.6 \times 10^{-6}$ c/d.  We preceeded the computation of
the final, high-resolution spectrum with a low-resolution scan (at 
approximately the minimum frequency separation imposed by the span of
the dataset) to determine
the correct mean value of the data from the zero-frequency amplitude.  We 
heavily oversampled the final spectrum in the frequency domain by a factor of
64, and used a gain of 
0.05 per clean iteration to make the cleaning process as numerically stable 
as possible; larger gain values and lower frequency resolution produced nearly
identical frequencies at the cost of slightly higher noise, while the cost in
oversampling is simply increased computation time.  The final, cleaned spectrum
of the 6.5-year, $V$-band data set is shown in Figure \ref{fig-ftclean}.  
There are three groups of frequencies of interest:
the primary pulsation frequency, $f_{0}$, and its integer-multiple harmonics
$2f_{0}$ to $5f_{0}$; a secondary frequency, $f_{1}$; and a low-frequency
peak, $f_{\rm{low}}$, that appears to be a beat frequency of 
$f_{\rm{low}} \sim f_{1} - f_{0}$.  The frequencies, amplitudes, and phases
of these frequencies are given in Table \ref{freqtable}.

The spectrum is dominated by the primary pulsation frequency of 
$f_{0} = 5.789939 \times 10^{-2}$ c/d ($P_{0} = 17.27134$ d) and its integer
multiples.  The higher order harmonics have significant amplitude, and as
a result the light curve is non-sinusoidal.  We compared the $V$-band Fourier 
component parameters $\phi_{j1} = \phi_{j} - j\phi_{1}$ and 
$R_{j1} = a_{j}/a_{1}$ of these integer multiple frequencies with previously 
published values \citep{Morgan03},
and find that our values are reasonably consistent with prior observations.
We note that our values were derived directly from the cleaned Fourier
spectrum; these values are often derived via sine wave fitting, and are
done to lower order, and so may not be strictly comparable.
A comparison of our derived parameters and those
published elsewhere is given in Table \ref{fourtable}.  Our derived values
of $R_{j1},\phi_{j1}$ are consistent with previously published values obtained
from other observations, with the only minor exception being $R_{31}$ which 
is lower by a factor of two in our data.  As a whole, this shows that the
underlying pulsations driven at $P_{0}$ are stable throughout the recent
observational record, indicating that the underlying pulsational waveform
is unchanged over recent years.  This is good evidence
for W Vir being a stable pulsator despite it being a long-period Cepheid.
Like the other long-period type II Cepheids (and unlike the BL Her stars and
classical Cepheids), there is no clear trend in $R_{j1},\phi_{j1}$ with
period.  However, the relatively large value of $R_{21} \sim 0.2$ is quite
similar to those of other high-amplitude pulsators, while the value of 
$\phi_{21} \sim 5.9$ (very close to $2\pi$) results in a nearly symmetric
(if non-sinusoidal) light curve unlike the sawtooth shapes of shorter
period Cepheids where $\phi_{21} \sim 4.0$.

\subsection{Updated ephemeris and $(O-C)$ analysis}

The $V$-band data set is of sufficient length and photometric quality that
we are able to compute an accurate, current ephemeris for W Vir.  We can then
compute an $(O-C)$ curve, and establish a rate of period change based upon
these archival data.  To determine the current ephemeris, we created a 
synthetic light curve using the frequencies, amplitudes, and phases for
$f_{0}$ through $5f_{0}$ given in Table \ref{freqtable}, excluding $f_{1}$
and $f_{low}$.  After ensuring that this curve was a good fit to the $V$-band
light curve, we chose the best-observed maximum (HJD 2452758.172) nearest the 
temporal center of the data set (HJD 2453078.271) as the zero-point of our 
ephemeris:

\begin{equation}
{\rm{HJD_{max}}} = 2452758.172 + 17.27134 \times {\rm{E}}
\end{equation}

We then used this ephemeris to measure $(O-C)$ for all available data for
W Vir.  The available data includes visual observations from the AAVSO
{\it International Database} spanning JD 2419500 to the present, along with
previously published photometry of various kinds collected by L. Berdnikov
and collaborators (Berdnikov 2002, private communication).  When the data
were sufficient to allow fitting, individual cycles were fit by eye using a 
synthetic mean curve placed at
the closest predicted time of maximum.  We adjusted both the magnitudes and
times of the observed data points to obtain a best fit to the synthetic 
cycle; a magnitude adjustment was often required due both to the difference in
response between visual estimates and $V$-band, and to cycle-to-cycle
variations not included in the synthetic mean curve.  For cases where there
was no objective, unambiguous way to improve the fit with either a magnitude 
or a temporal
offset, we left one or both equal to zero; for this reason, many of the
$(O-C)$ measures are ``0.0'', skewing the $(O-C)$ diagram close to the zero
point of the ephemeris.  For the visual data, we used individual 
observers when possible to minimize scatter between the different subjective
responses by different observers, but the $(O-C)$ values obtained from visual 
data were marginal in general due to incomplete coverage of pulsation 
cycles.  Although AAVSO data prior to 1932 exist, they are generally
not of sufficient quantity or homogeneity to allow unambiguous cycle fitting.  
Fits to instrumental photometry were generally better
though they too suffer from incomplete coverage of cycles, and are also 
affected by the intrinsic cycle-to-cycle variations.  We initially estimated
errors on the times of maxima to be around 0.5 days, and the resulting
reduced $\chi^{2}$ values of the fits suggest the errors are likely around
0.35 to 0.4 days.  

\begin{figure}
\figurenum{3}
\label{figo-c}
\epsscale{0.75}
\plotone{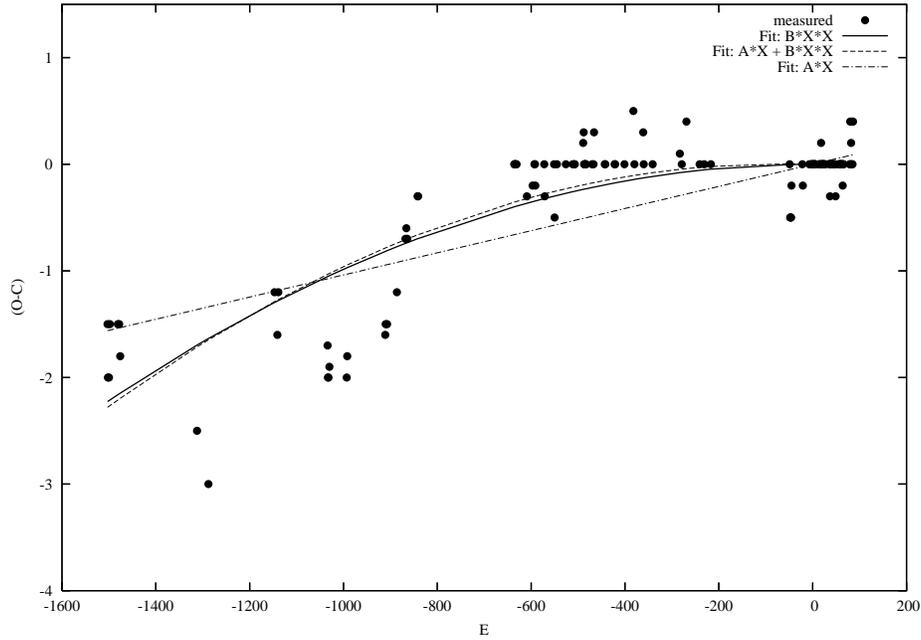}
\caption{$(O-C)$ diagram of W Vir using the ephemeris given in equation 1.
There is substantial variation in $(O-C)$ with each cycle, but the overall
trend is parabolic which indicates a secular period change.  The reasons for
the cycle-to-cycle scatter are: (1) that the seconary pulsation frequencies 
affect the time of maximum and the goodness of the fit, and (2) that the 
sparse data for most of the cycles prior to the {\it ASAS-3} data make it
very difficult to fit a mean curve reliably.  The approximate error in
$(O-C)$ measures is 0.3 days.
}
\end{figure}

The resulting $(O-C)$ diagram for W Vir is shown in Figure \ref{figo-c},
with first- and second-order polynomial fits superimposed.  There is
clearly a significant trend in the $(O-C)$ curve, and the period-change
term is identical to within error bars for fits involving only a second-order
term, and a quadratic (first and second order term) fit.  The reduced 
$\chi^{2}$ value for the pure second-order fit is the same as that for the 
quadratic fit.  A
fit involving only a first-order term (which measures only an error in the
period, rather than a period change) has a worse $\chi^{2}$ value than either of
the other two fits.  Because of this, we are confident that we have detected
the period change of W Vir over the span of data used in this paper.  The 
resulting ephemeris with a second-order correction is

\begin{equation}
{\rm{HJD_{max}}} = 2452758.172 + 17.27134 \times {\rm{E}} - 9.9 \times 10^{-7} \times {\rm{E}}^{2}
\end{equation}

Our result on the last 75 years of data contradicts the inconclusive results
found by \citet{PH00}, who found essentially no period change when analyzing
archival times of maximum and minimum dating to the 19th Century.  We have
not included data prior to 1932 in our analysis, but we note the existence of
both sparse AAVSO observations and published photometry along with archived 
times of maximum and minimum, and plan to reanalyze these data for a future
paper.

\subsection{BVRcIc light curves}

\begin{figure}
\figurenum{4}
\label{fig-sonlc}
\epsscale{0.75}
\plotone{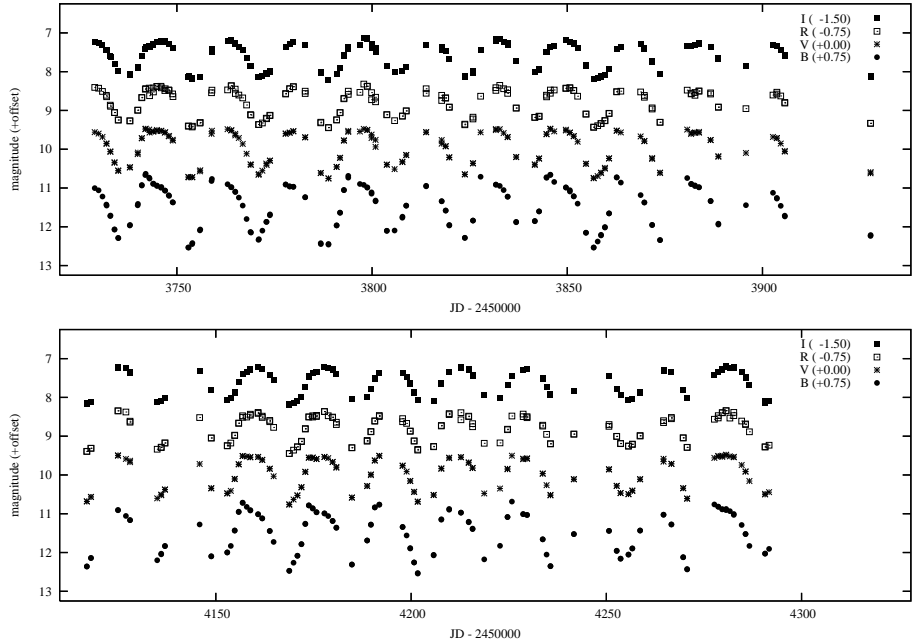}
\caption{BVRcIc light curve.  The following magnitude offsets were applied to
the $B$, $Rc$, and $Ic$ data for clarity: $B$ (+0.75); $Rc$ (-0.75); $Ic$ (-1.5).
}
\end{figure}

\begin{figure}
\figurenum{5}
\label{fig-sonfold1}
\epsscale{0.75}
\plotone{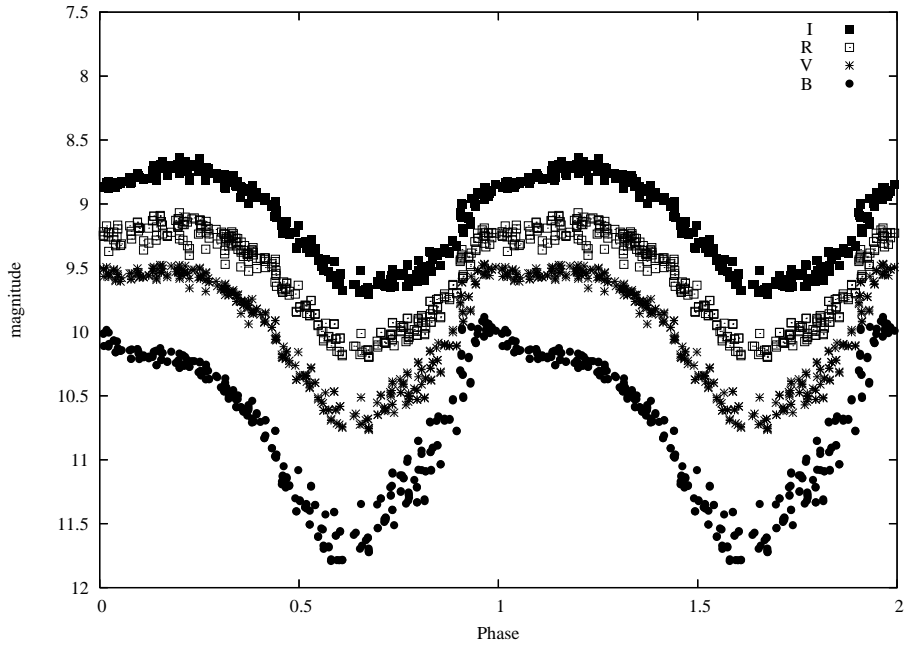}
\caption{The $BVRcIc$ light curve folded on the fundamental period of
17.27134 days.
}
\end{figure}

A major part of this project was to observe W Virginis in $BVRcIc$ 
simultaneously to better understand the underlying physical behavior of
the pulsations, and provide full photometric coverage for the spectroscopy
performed by Wallerstein et al.  Very little multicolor work has been done
on W Vir since the early photometric studies performed in the 1950s.  Some
multicolor amateur data exists, with the most notable being the $BV$ 
photometry of M. Bonnardeau taken during the spring and summer of 2005,
and the AAVSO data submitted as part of our support for the Wallerstein
campaign.  The observations we obtained with SRO represent the largest
multicolor data set obtained for W Vir in its history.  The full $BVRcIc$ 
light curve taken at SRO is shown in Figure \ref{fig-sonlc}, and the $BVRcIc$ 
light curve folded on a period of 17.27134 days is shown in Figure
\ref{fig-sonfold1}.  We note that these light curves include all data through
the end of the Spring 2007 observing season (JD 2454299), which were obtained 
and reduced after the bulk of the analysis for data through JD 2454255 was 
completed for this paper.  The additional 40 days of data do not change the 
main results given in the prior sections of this work and are included to 
provide more complete phase coverage of the 2007 season.  

\subsubsection{Color-dependencies of the pulsation behavior}
\begin{figure}
\figurenum{6}
\label{fig-rval}
\epsscale{0.75}
\plotone{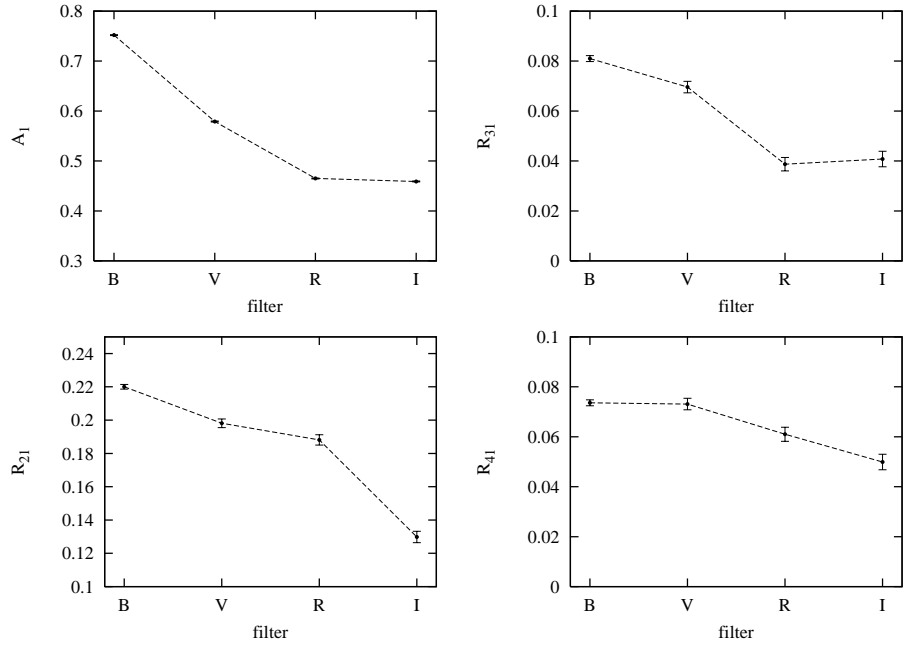}
\caption{$BVRcIc$ amplitude ratios of Fourier harmonics.
}
\end{figure}

\begin{figure}
\figurenum{7}
\label{fig-dphi}
\epsscale{0.75}
\plotone{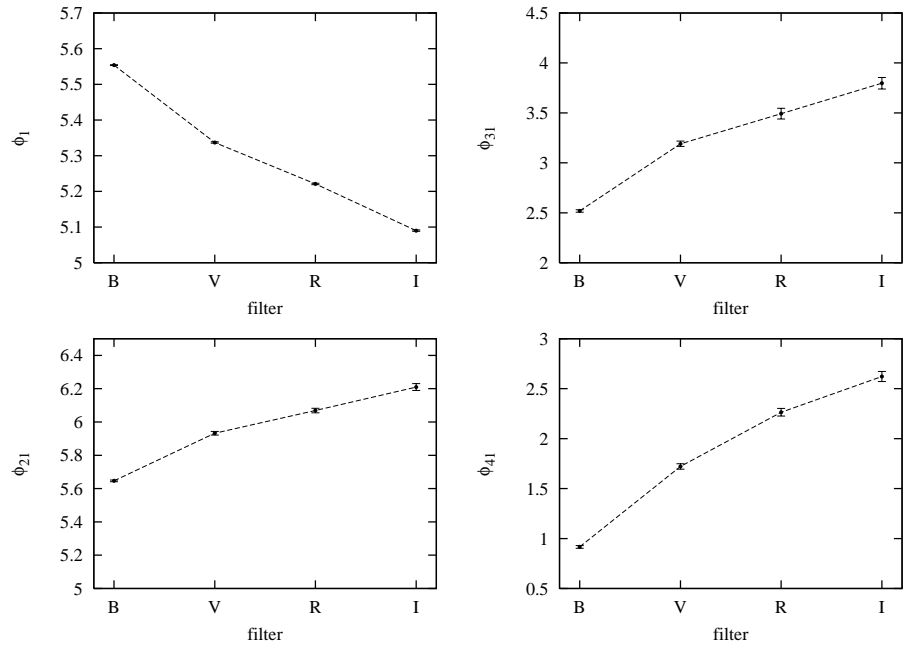}
\caption{$BVRcIc$ phase differences of Fourier harmonics.
}
\end{figure}

Pulsation amplitude is highest in $B$ and weakens progressively through $V$,
$Rc$, and $Ic$.  The $B$-band light curve has a sharp peak followed by a short
plateau and decline, while the other three bands rapidly rise to a
plateau and remain constant for approximately half a cycle.  Maximum
light occurs immediately following the rapid rise at $B$ and $V$, while
$Rc$ remains essentially flat, and $Ic$ has a slowly rising plateau with
peak approximately 0.2 cycles after $B$ and $V$. We determined the
Fourier amplitude ratios $R_{j1}=a_{j}/a_{1}$ and phase differences
$\phi_{j1} = \phi_{j} - j\phi_{1}$ for $j = 2,4$ for each of the 
$BVRcIc$ light curves.  The
amplitude ratios are shown in Figure \ref{fig-rval}, and the phase
differences are shown in Figure \ref{fig-dphi}.  The amplitude ratios are 
largest in $B$ and smallest in $Ic$ for all harmonics, indicating less
sinusoidal light variations at bluer wavelengths.  The value of $\phi_{21}$ 
is close to $2\pi$ in all four bands, which results in a much more 
symmetric light curve, and it asymptotically approaches $2\pi$ at 
redder wavelengths.  The values of $\phi_{31}$ and $\phi_{41}$ would produce 
larger asymmetries, but the harmonics have lower amplitude and thus a smaller
effect on the overall light curve.  Generally, the amplitude ratios and 
phase differences indicate that the light curves become increasingly 
symmetrical (though not purely sinusoidal) at redder wavelengths.

The $(B-V)$ color spans a much larger range (+0.4 at maximum to +1.1 at
minimum) than do $(V-Rc)$ or $(Rc-Ic)$ because the temperature range is large
($\sim 5000-7000$ K) and the spectral peak at $T_{\rm max}$ lies within
the $B$ band.  The color change is very rapid during rising light, going
from reddest to bluest within 0.2 cycles; the steepness of both the 
light curve and the color curve suggest a strong shock, and this is consistent
with past spectroscopic studies (see \citet{LG92} and references therein).

\subsubsection{Cycle-to-cycle variations}
\begin{figure}
\figurenum{8}
\label{fig-cc}
\epsscale{0.85}
\plotone{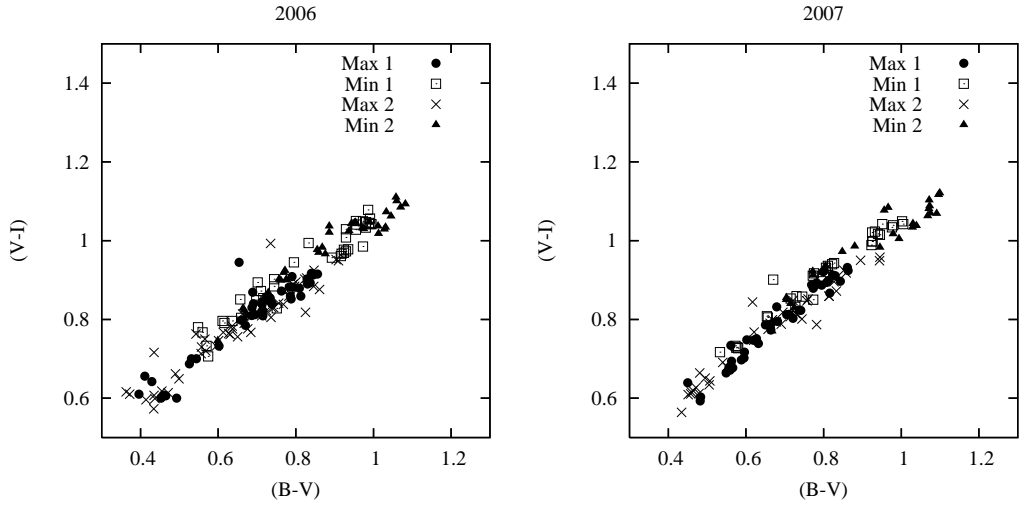}
\caption{$(B-V)$ versus $(V-Ic)$ plots for the 2006 and 2007 data sets, where
the data are folded with $2P_{0}$, and the resulting two maxima and minima
are plotted separately.  Maxima and minima are defined by $V < 10$ and $V > 10$
respectively.  In both years, the second minima is significantly redder in
$(B-V)$ than the first minima, indicating that there is a physical difference
between the two, and that the minima are alternating in some way.
}
\end{figure}

The folded light curve in Figure \ref{fig-sonfold1} shows much larger scatter
($> 0.1$ mag) than the photometric errors (generally $<0.01$ mag), indicating
that there are cycle-to-cycle variations.  This is expected given the
amplitudes of $f_{\rm low}$ and $f_{1}$; folding on twice the fundamental
period decreases the scatter considerably, although it does not remove 
it entirely, particularly for the 2007 season.  Color-color plots of
$(B-V)$ versus $(V-Ic)$ for single-season data folded on twice the fundamental
period are shown in Figure \ref{fig-cc}.  The color-color plot clearly shows
that the deeper of the two minima (Min 2) has a significantly redder $(B-V)$
color than (Min 1), indicating a cooler temperature.  The fact that this
difference between the two minima persists throughout the season indicates
that the pulsations are being modulated with a period approximately twice
that of the main period.

\subsubsection{$Rc$-band irregularity}

The $Rc$-band light curve shows much larger scatter than $B$, $V$, or $Ic$
when folded, indicating that there is a transient, irregular spectral
feature occurring within the $Rc$ band.  We have no means of uncovering
details of this feature with broadband photometry, but the work of \citet{LG92}
hints strongly that it may be variable post-shock emission lines in an
extended atmosphere.  \citet{LG92} found very large variations in H$\alpha$
flux over the course of the pulsation cycle, as well as irregularity in
the evolution of the line profile variations.  The flux variation of the
H$\alpha$ emission alone is about five percent; the width of the line is
about 0.5\% of the width of the $Rc$ band, and the flux varies between
0.5 and 6 times the continuum, which could generate variations at the level
of about 0.1 magnitudes.  But what is striking is that there are large
variations at a given phase from cycle to cycle, which indicates that the
shock features are not occurring at the same time as they propagate into
the extended atmosphere, and do not have the same strength.  This suggests
that the response of the extended atmosphere is irregular, despite the
regularity of the photospheric motions.  High-resolution spectroscopy 
over several cycles may provide more insight into this behavior.

\section{Discussion}

There are several important results of our work.  We conclusively show
that there are cycle-to-cycle variations in the light curve of W Vir, but 
that these variations appear to be due to the presence of two pulsation
modes, and not to inherent instabilities in the pulsation.  The fact that 
these modes are well-defined in the Fourier spectrum suggests that the 
pulsation behavior is largely linear in nature.  Prewhitening of the 
$V$-band data with the frequency set in Table \ref{freqtable} does not fully
remove the signal, which suggests that a second process is at work -- either 
a secular period change or low-level nonlinearity.  The period of W Vir is 
definitely changing, as was shown by the $(O-C)$ diagram in Figure 
\ref{figo-c}, and the fact that the prewhitening is clearly worse at the 
temporal extremes of the light curve than at the temporal center suggests 
that a secular period change may explain long-term changes without 
requiring nonlinear behavior.

The secondary pulsation frequencies $f_{\rm{low}}$ and $f_{1}$ are real, 
and are not aliases of the main frequency.  The color differences between 
alternating minima seen in the 2006 $BVRcIc$ data must have a physical 
origin, and a beat frequency of $f_{\rm{low}} = f_{1} - f_{0} \sim f_{0}/2$ 
is the easiest interpretation.  The fact that the pattern of alternating 
minima did not repeat as clearly during the 2007 observing season is caused 
by the low-frequency peak $f_{\rm{low}}$ not being exactly $f_{0}/2$.  The
two secondary spectral peaks could be a manifestation of {\it period-doubling},
as was described (for example) in \citet{Pollard00}.  However, two facts 
suggest a purely modal explanation for W Vir rather than a fundamental change 
in pulsation
behavior.  First, the two peaks in question are not {\it exact} rational 
fractions of the main frequency to within the frequency precision of either
peak.  Any nonlinear process which ``doubles'' the primary period without
making it exactly twice the period while keeping it relatively stable over
time would be difficult to explain.  Second, the frequency ratio
$f_{0}/f_{1} \sim 0.667$ is consistent with the likely frequency ratio of 
the radial fundamental and first overtone modes of Population II
Cepheids having periods longer than 10 days.  (See \citet{Buchler07}
for recent models.)  

We propose that the cycle-to-cycle variations of W Vir
are best explained in terms of combinations of linear pulsation modes, and
that the strength of these variations is due to the near-resonant ratio of
the fundamental and first overtone frequencies.
We note that because this particular frequency ratio should only occur in
well-evolved, long-period Cepheids (possibly those on their last crossing),
this stage may be the last waypoint in the evolution of W Vir stars before
they move up the AGB into the RV Tauri phase.
For the general population of long-period Type II Cepheids that show strong
cycle-to-cycle variability, their apparent irregularity may be due to overtone 
modes instead of nonlinear pulsation behavior.

The link between the W Virginis stars and the RV Tauri stars has been
discussed previously (see \citet{Wallerstein02} for example), and 
observational work on the LMC W Virginis and RV Tauri sample 
\citep{Pollard00} shows a progression in the physical and pulsational
characteristics of these two classes.  The physical distinction between the
two in this scenario is that the W Vir stars are slightly less evolved members
of the same population whose lower $L/M$ ratios keep them within the linear
pulsation regime; the RV Tauri stars, having larger $L/M$ ratios, are in the
nonlinear regime.  Interestingly, although RV Tauri
stars are clearly nonlinear in nature, the work of \citet{Buchler96}
showed that their seemingly chaotic behavior is confined to a low-dimensional
phase space, and they interpret this as being due to the 
interaction of as few as two pulsation modes -- exactly what we see with
W Vir.

The picture that emerges is that double- or multiperiodicity becomes 
increasingly likely at longer periods and larger $L/M$ ratios among
Population II giants, and that both the Type II Cepheids of long 
period and the RV Tauri stars exhibit this
behavior.  The phenomenological difference between the two is then simply
that the RV Tauri stars have reached the point where nonlinearities overwhelm
the pulsational dynamics of the envelope, leading to irregularity in the
light curve.  We suggest three future studies that could confirm this picture.
First, evolution and nonadiabatic, linear pulsation models of W Vir stars 
having periods longer than 10 days are crucial to studying the physical link
between the two classes, and we strongly encourage the Cepheid modeling
community to expand their model grids to these longer periods and higher
luminosities into the RV Tauri region.  Second, a long-term photometric 
study of other
long-period Type II Cepheids should be undertaken to search for linear
multiperiodicities.  The existence of many all-sky photometric surveys such
as {\it ASAS} \citep{Pojmanski02} greatly facilitates this, and such an
analysis is currently underway (Wils \& Otero, private communication).
Finally, an analysis of the chaotic dynamics of a larger sample of RV Tauri
stars (and the Type II Cepheids) should be done, as was performed on R Scuti
\citep{Buchler96} and AC Her \citep{Kollath98}.  Such studies could easily be
performed on both large-scale photometric survey databases, and on existing
archives of long-term visual observations.

\section{Conclusions}

Despite being the class prototype of a cosmologically important class of
variable stars, our understanding of W Virginis and other Type II Cepheids
is incomplete.  Until now, little concentrated observational work has been
done on this object other than with single-cycle photometry or sparse coverage
over a single observing season with photometry and spectroscopy.  The use of
an automated telescope with a well-planned observing queue made it possible to
follow W Virginis regularly for nearly two years.  This highlights the power
of small, automated telescopes for performing long-term studies of neglected
stars, and also the fact that there remains work still to be done even on
bright, ``well-studied'' objects.

Our work has clarified some long standing questions about
the pulsation behavior, and confirmed the existence of double-mode pulsations
and the resulting cycle variations in this star.  Despite the existence of
a second pulsation period, the pulsational signature of the dominant mode
is very stable over time, suggesting that W Vir at least does not show
incipient instabilities like those seen in some long-period Type II Cepheids.
An interesting question then becomes where is the boundary between the
parameter spaces of the Type II Cepheids and the RV Tauri stars, and at what
point to the linear pulsations of Cepheids give way to the nonlinear behavior
shown in the RV Tauri stars?  Linear, nonadiabatic pulsation models of both 
Type II Cepheids and RV Tauri stars clearly need to be generated using modern
opacities to map out the parameter space of mass, luminosity and temperature
to determine fundamental and overtone mode periods and their nonadiabatic 
growth rates; hydrodynamic models will be key to understanding where the
transition between linearity and nonlinearity lies, and we encourage the
computation of both by the theoretical modeling community.

\acknowledgements
We wish to thank J.~Gross, D.~Terrell, and W.~Cooney for their permission to
use the {\it Sonoita Research Observatory} for this project.  As always,
we acknowledge with gratitude the observations made by members of the AAVSO's
worldwide observing community over the last century, without whom this work
would not be possible.  We also thank L.~Berdnikov for supplying us with the
compiled published photometry in 2002.  And we thank the anonymous referee
whose comments helped to clarify and improve the paper.  This research has 
made use of NASA's Astrophysics Data System, and the SIMBAD database, operated
at CDS, Strasbourg, France.

\clearpage

%\eject

\clearpage

\begin{deluxetable}{rrrrrrrrrrrr}
\rotate
\tablenum{1}
\tablecolumns{12}
\tablewidth{0pc}
\tablecaption{W Vir Comparison Stars}
\tablehead{\colhead{RA(J2000)} & \colhead{DEC(J2000)} & \colhead{V} & 
\colhead{(B-V)} & \colhead{(V-Rc)} & \colhead{(Rc-Ic)} & \colhead{(V-Ic)} & 
\colhead{err(V)} & \colhead{err(B-V)} & \colhead{err(V-Rc)} & \colhead{err(Rc-Ic)} & 
\colhead{err(V-Ic)}} 
\startdata
201.395994 & -3.296183 & 13.051 & 0.811 & 0.436 & 0.387 & 0.820 & 0.006 & 0.008 & 0.005 & 0.004 & 0.005 \\201.292630 & -3.312171 & 13.564 & 0.970 & 0.568 & 0.544 & 1.112 & 0.005 & 0.009 & 0.005 & 0.005 & 0.005 \\
201.244004 & -3.399964 & 12.174 & 0.581 & 0.351 & 0.342 & 0.715 & 0.006 & 0.009 & 0.005 & 0.005 & 0.007 \\
201.264303 & -3.546543 & 11.043 & 1.110 & 0.569 & 0.494 & 1.058 & 0.005 & 0.008 & 0.004 & 0.004 & 0.005 \\
201.264437 & -3.528959 & 13.328 & 0.972 & 0.560 & 0.467 & 1.020 & 0.005 & 0.007 & 0.004 & 0.004 & 0.005 \\
201.351088 & -3.528159 & 12.517 & 0.564 & 0.323 & 0.294 & 0.615 & 0.005 & 0.008 & 0.004 & 0.004 & 0.005
\enddata
\label{compstars}
\end{deluxetable}

\begin{deluxetable}{crccrcc}
\tablenum{2}
\tablecolumns{7}
\tablewidth{0pc}
\tablecaption{Top seven peaks in the cleaned Fourier spectrum of W Vir.  The
Fourier amplitudes and phases of the harmonics $2f_{0}$ through $5f_{0}$
are given at the exact frequency, rather than at the measured peak, because
background noise shifts the location of the peak slightly from the true value.
}
\tablehead{\colhead{frequency} & \colhead{freq. error} & 
\colhead{amplitude} & \colhead{amp. error} & 
\colhead{phase} & \colhead{phase error} & \colhead{ID} \\
\colhead{(c/d)} & \colhead{($\times 10^{-8}$)} &
\colhead{(V mag)} & \colhead{(V mag)} &
\colhead{(rad)} & \colhead{(rad)} & \colhead{}
}
\startdata
 0.05789939 & 2   & 0.590 & 0.001 & -0.716 & 0.001 & $f_{0}$\\
 0.11579878 & 12  & 0.119 & 0.001 & -1.820 & 0.006 & $2f_{0}$\\
 0.17369817 & 43  & 0.033 & 0.001 &  1.052 & 0.021 & $3f_{0}$\\
 0.23159755 & 27  & 0.052 & 0.001 & -1.227 & 0.016 & $4f_{0}$\\
 0.28949694 & 157 & 0.009 & 0.001 &  2.543 & 0.059 & $5f_{0}$\\
 0.02890449 & 32  & 0.044 & 0.001 &  1.414 & 0.035 & $f_{\rm{low}}$\\
 0.08676324 & 62  & 0.023 & 0.001 &  2.791 & 0.054 & $f_{1}$
\enddata
\tablecomments{Phases are relative to HJD = 2453078.271, the time
center of the {\it ASAS-3 + SRO} data set.}
\label{freqtable}
\end{deluxetable}

\begin{deluxetable}{llllllllllc}
\rotate
\tablenum{3}
\tablecolumns{11}
\tablewidth{0pc}
\tablecaption{Fourier amplitude ratios $R_{j1}$ and phase differences 
$\phi_{j1}$ for W Vir.  Previously published measures were taken from
\citet{Morgan03}.}
\tablehead{\colhead{Period} & \colhead{$A_{1}$} & \colhead{$R_{21}$} & 
\colhead{$R_{31}$} & \colhead{$R_{41}$} & \colhead{$R_{51}$}
& \colhead{$\phi_{21}$} & \colhead{$\phi_{31}$}& \colhead{$\phi_{41}$}& 
\colhead{$\phi_{51}$} & \colhead{Source}
\\
\colhead{(d)} & \colhead{(mag)} & \colhead{} & \colhead{} &
\colhead{} & \colhead{} & \colhead{(rad)} & \colhead{(rad)} & \colhead{(rad)}
& \colhead{(rad)} & \colhead{}
}
\startdata
17.27134 & 0.591 & 0.201 & 0.056 & 0.089 & 0.015 & 5.896 & 3.200 & 7.920 & 
6.124 & (this study)\\
17.267 & 0.6453 & 0.2164 & 0.1040 & 0.1084 & ... & 5.7878 & 3.2716 & 8.6234 &
 ... & 1 \\
17.274 & ... & ... & ... & ... & ... & 5.9100 & 3.2168 & 8.1800 & 
 ... & 2 \\
17.3 & 0.6130 & 0.1480 & 0.0940 & 0.0740 & ... & 6.080 & 2.8368 & 8.1700 &
 ... & 3 
\enddata
\tablecomments{Sources: (1) -- \citet{Zakrzewski00}, $V_{\rm{Hip}}$;
(2) -- \citet{Fernie99}, $V$;
(3) -- \citet{Simon93}, $V$
}
\label{fourtable}
\end{deluxetable}

\end{document}